\documentclass[preprint,nofootinbib,aps,superscriptaddress,floatfix]{revtex4}

\usepackage[dvips]{graphics,color}

\begin{document}

\def\LQCD{\Lambda_{\rm QCD}}
\def\lqcd{\Lambda_{\rm QCD}}
\def\xslash#1{{\rlap{$#1$}/}}
\def\dsl{\,\raise.15ex\hbox{/}\mkern-13.5mu D}
\def\half{{\textstyle{1\over 2}}}
\def\halfpi{{\textstyle{\pi\over 2}}}
\def\ndotq{n \!\cdot \! q}
\def\nbardotq{\bar{n} \!\cdot \! q}
\def\ndotk{n \!\cdot \! k}
\def\nbardotk{\bar{n} \!\cdot \! k}
\def\ndotl{n \!\cdot \!\ell}
\def\ndots{n \!\cdot \! s}
\def\nbardotl{\bar{n} \!\cdot \!\ell}
\def\lone{\ell_1}
\def\ndotlone{n \!\cdot \!\lone}
\def\ndotD{n \!\cdot \! D}
\def\btou{\bar{B} \to X_u \ell \bar{\nu}_\ell}
\def\shhat{\hat s_H}
\def\ltapp{\lesssim}
\newcommand{\Budecay}{$\bar{B} \to X_u \,  \ell \,  \bar{\nu} $ }
\newcommand{\Bcdecay}{$ \bar{B} \to X_c \,  \ell \,  \bar{\nu} $ }
\newcommand{\Bsgamma}{$ \bar{B} \to X_s \,  \gamma $ }
\newcommand{\orderalpha}{ {\cal{O}}(\alpha_s)}
\newcommand{\btoc}{\bar{B} \rightarrow \rm X_c \, \ell \, \bar{\nu}}


\def\ctp#1#2#3{\CTP{\bf #1} (#2) #3}
\def\jetpl#1#2#3{\JETPL{\bf #1} (#2) #3}
\def\nc#1#2#3{\NC{\bf #1} (#2) #3}
\def\np#1#2#3{\NP{\bf B#1} (#2) #3}
\def\pl#1#2#3{\PL B {\bf #1} (#2) #3}
\def\prl#1#2#3{\PRL{\bf #1} (#2) #3}
\def\prd#1#2#3{\PR D {\bf #1} (#2) #3}
\def\prep#1#2#3{\PRep{\bf #1} (#2) #3}
\def\physrev#1#2#3{\PR{\bf #1} (#2) #3}
\def\sjnp#1#2#3{\SJNP{\bf #1} (#2) #3}
\def\nuvc#1#2#3{\NC{\bf #1A} (#2) #3}
\def\blankref#1#2#3{   {\bf #1} (#2) #3}
\def\ibid#1#2#3{{\it ibid,\/}  {\bf #1} (#2) #3}
\def\AP{{\it Ann.\ Phys.\ }}
\def\CMP{{\it Comm.\ Math.\ Phys.\ }}
\def\CTP{{\it Comm.\ Theor.\ Phys.\ }}
\def\IJMP{{\it Int.\ Jour.\ Mod.\ Phys.\ }}
\def\JETPL{{JETP Lett.\ }}
\def\NC{{\it Nuovo Cimento\ }}
\def\NP{{Nucl.\ Phys.\ }}
\def\PL{{Phys.\ Lett.\ }}
\def\PR{{Phys.\ Rev.\ }}
\def\PRep{{Phys.\ Rep.\ }}
\def\PRL{{Phys.\ Rev.\ Lett.\ }}

\newcommand{\nn}{\nonumber}

\title{Towards the Anomalous Dimension to $\mathcal{O} (\LQCD/m_b)$ for Phase Space Restricted  \Budecay and  \Bsgamma }

\author{Michael Trott }\email{mrtrott@physics.ucsd.edu}

\affiliation{Department of Physics, University of California at San Diego,\\[-5pt]
  La Jolla, CA, 92093}

\author{Alexander R. Williamson}\email{alexwill@andrew.cmu.edu}

\affiliation{Department of Physics, Carnegie Mellon University,\\[-5pt]
 Pittsburgh, PA, USA, 15213}

\begin{abstract}
We examine the anomalous dimension matrix appropriate for the phase space restricted  \Budecay and  \Bsgamma decay spectra to subleading nonperturbative order. 
The time  ordered products of the HQET Lagrangian with the leading order shape function operator are calculated, 
as are the anomalous dimensions of subleading operators. 
We establish the renormalizability and closure of a subset of the non-local operator basis, a requirement for the establishment of factorization theorems at this order. Operator mixing is found between the operators which occur to subleading order, requiring the subleading operator basis be extended.  We comment on the requirement for new shape functions to be introduced to characterize the matrix elements of these new operators, and the phenomenological consequences for extractions of $|V_{ub}|$. 
\end{abstract}

\maketitle

\section{Introduction}
Extracting the CKM parameter $|V_{ub}|$ is an important
step in testing of the CKM description of CP violation in the $B$ meson system. 
Currently, the theoretically cleanest determinations of  $|V_{ub}|$ 
come from inclusive semileptonic decays which are not sensitive to the details 
of hadronization;  although recently  an approach of extracting $|V_{ub}|$, utilizing 
$B \rightarrow \pi \, \pi$, has been advanced with a competitive error to inclusive methods. \cite{hep/0504209}

In inclusive extractions of $|V_{ub}|$, experimental cuts to exclude the charm background of \Bcdecay 
are imposed. This restricts the decay products to hadronic 
final states that have large energy $E_X \sim m_B$ and low invariant mass 
$ M_X \sim \sqrt{m_B \lqcd}$. With these 
phase space restrictions the local OPE expansion \cite{operefs} appropriate for sufficiently 
inclusive decays used to extract  $|V_{cb}|$ \cite{BLLMT2004},  typically breaks down. \cite{Luke0307378}

As the local OPE, and the clean separation of scales that the local OPE represented in the analysis of  \Bcdecay, is no longer valid; 
a more involved theoretical approach is required to separate the scales relevant to these decays. 
Decay rates are expressed as convolutions of hard ($H$), jet ($J$) and soft physics ($S$) associated 
with the scales $m_b \gg \sqrt{\LQCD \, m_b} \gg \LQCD $, in the following way,
\begin{eqnarray} \label{convolution}
{\rm d} \, \Gamma =  H \left(\frac{m_b}{\mu}, \alpha_s(\mu) \right) \, \int  d \, \omega \, J \left(\frac{\sqrt{m_b \, \LQCD}}{\mu}, \alpha_s(\mu),\omega \right)  \, S(\omega).
\end{eqnarray}
Although this factorization theorem has been proven diagrammatically \cite{Sterman} at leading order in $1/m_b$, it is not known to hold to all orders in the nonperturbative expansion. 
It has only recently been extended beyond leading nonperturbative order \cite{LeeStewart}. 

The systematic treatment of the nonperturbative corrections involve a two step
matching procedure. One matches QCD onto the effective field theory of the intermediate scale, 
describing quarks and gluons with large energy and small offshellness, known as SCET \cite{SCETlukebauer01,SCETall01,SCETall02},
and uses the renormalization group evolution to run down to the soft scale.
One then matches SCET onto the lightcone wavefunction of the $B$ meson, expressed in terms of HQET fields.
One can also match directly from QCD onto HQET, a much
simpler procedure at the cost of not summing the logarithms of the ratio of scales
$\log \left( \sqrt{  m_b \lqcd} /  m_b \right) $ via SCET. 
In either case,  the soft sector of the theory is expanded in terms of non-local operators.
The leading order term in the $\LQCD/m_b$ expansion of  
the lightcone distribution function of the $B$ meson \cite{shapeleading1,shapeleading2} 
is know as the shape function.

At subleading order in the nonperturbative expansion, 
four additional non-local  operators have been determined to be present \cite{blm01, blm02, mn02, llw02, Bosch:2004cb}, the matrix elements of which are referred to as subleading shape functions.

It is of some intrinsic interest to examine the renormalization of these non-local operators, as they are non-local and their renormalizability is not know {\it a priori}. 
It is also important to know if this set of operators is complete in the error assigned to extractions of $|V_{ub}|$, and in considering the above factorization theorem beyond leading order. Examining the perturbative behavior of these subleading shape functions is also a necessary step to take to perform one loop matching calculations onto the soft sector. 
For these reasons we have determined the anomalous dimension to subleading nonperturbative order.

To this end, we have determined the contributions of the time ordered products of the subleading  $\cal{L}_{\rm HQET} $ with the leading order shape function, and examined the anomalous dimensions of the subleading nonperturbative operators. We establish the renormalizability and closure of a subset of the subleading non-local operators. We find that the known operator basis mixes with new operators, requiring that the subleading operator basis be extended. We also comment on the phenomenological consequences of these results.

\section{Anomalous Dimension to Subleading Order}

\subsection{Notation}

We introduce two light-like vectors
$n^\mu$ and $\bar{n}^\mu$ related to the velocity of the heavy quark
by $v = \half(n+\bar{n})$, and satisfying
\begin{equation}
n^2 = \bar{n}^2 = 0, \quad v\cdot n = v\cdot \bar{n} = 1, \quad n\cdot
\bar{n} = 2.
\end{equation}
In the frame in which the $b$ quark is at rest,
these vectors are given by $n^\mu=(1,0,0,1)$, $\bar
n^\mu=(1,0,0,-1)$ and $v^\mu=(1,0,0,0)$. The projection of an arbitrary 
four-vector $a^\alpha$ onto the directions which are perpendicular to 
the lightcone is given by
$ a_\perp^\alpha = g_{\perp}^{\alpha\beta} a_\beta$, where
\begin{eqnarray}
g^{\mu \nu}_\perp &\equiv& g^{\mu \nu}-\frac{1}{2} \left(n^\mu \bar{n}^\nu
+ n^\nu \bar{n}^\mu \right).
\end{eqnarray}
We also define a perpendicular Levi-Civita tensor
\begin{equation}
\epsilon_\perp^{\alpha\beta} = \epsilon^{\alpha \beta \sigma \rho} v_\sigma n_\rho.
\end{equation}
We also use the projector
$P_+ = 1/2(1+\xslash{v})$ as well as the Dirac structure
$s^\eta = P_+ \gamma^\eta \gamma_5 P_+$, so that $v\cdot s = 0$.

\subsubsection{Distribution Notation}
\label{Star Distribution Notation}

Rather than the usual definitions of the star distribution as given in Neubert and deFazio \cite{deFazioNeubert},
\begin{eqnarray}
\left(\frac{1}{x}\right)_*&=&\lim_{\beta\to0}
\left[\frac{\theta(x-\beta)}{x}+\delta(x-\beta)\log(x)\right]\nonumber\\
\left(\frac{\log(x)}{x}\right)_*&=&\lim_{\beta\to0}
\left[\frac{\theta(x-\beta)}{x}\log(x)+
\frac{1}{2}\delta(x-\beta)\log^2(x)\right],
\end{eqnarray}
we utilize the alternate notation, equivalent to the $\mu$-distribution's in \cite{Bauer:2003pi} 
\begin{equation}
\phi_n(x)\equiv\lim_{\beta\to 0}\left[\frac{1}{n+1}\theta(x-\beta)\log^{n+1}(x) \right].
\end{equation}
This notation has a fairly easy correspondence to the usual star distribution notation 
\begin{equation}
\phi'_0(x) =  \left(\frac{1}{x}\right)_*, \, \, \, \,
\phi'_1(x) =  \left(\frac{\log(x)}{x}\right)_*.
\end{equation}

A useful identity given by analytic continuation is
\begin{equation}\label{thetaeqn}
\frac{\theta(x)}{x^{1+\epsilon}} = -\frac{1}{\epsilon} \delta(x)+
\phi_0'(x) - \epsilon \phi_1'(x)
+ \mathcal{O}(\epsilon^2)
\end{equation}
This relationship is valid when integrated against arbitrary functions $f(x)$,
where $f$ is not singular at the origin.
In general we can write the recursion relation
\begin{equation}
\frac{\theta(x)}{x^{n+\epsilon}}=\frac{-1}{n\!-\!1\!+\!\epsilon}
\,\frac{d}{dx}\!\left[\frac{\theta(x)}{x^{n-1+\epsilon}}\right]\mbox{ for }n\ge2.
\end{equation}

Several other useful properties of this function are (for some positive constant $a$):
\begin{eqnarray}
x \, \phi_0'(x) &=& \theta(x) \nonumber \\
x \, \phi_0''(x) &=& \delta(x) - \phi_0'(x) \nonumber \\
a \, \phi_0'(ax) &=& \phi_0'(x) + \delta(x) \log(a),
\end{eqnarray}
as well as
\begin{eqnarray}
\int_{-\infty}^a f(x) \, \phi_0'(x) = \int_0^a \left(\frac{\theta(x)}{x} \right)_+  f(x) + f(0) \log(a).
\end{eqnarray}
For an $a$ with dimension, such as $a =\LQCD$, the above equation is suitably modified to obtain
logs the dimensionless ratio $a/ \mu$. 

\subsection{Operators to subleading Order}

At leading order a single non-local operator characterizes the nonperturbative physics,
\begin{equation} \label{leadingoperator}
Q_0(\omega,\Gamma) =  \bar{h}_v \, \delta(\omega + i \ndotD ) \, \Gamma \, h_v ,
\end{equation}
where the covariant derivative is $D_\mu = \partial_\mu + i g_s A_\mu$.

The order $\lqcd /m_b$ corrections to the \Budecay and  \Bsgamma decay spectra require the introduction of four additional non-local operators \cite{blm01, blm02, mn02, llw02, Bosch:2004cb},
\begin{eqnarray} \label{subleadingoperators}
m_b \,  Q_1^{\mu}(\omega,\Gamma)&=&
 \bar{h}_v \left\{i D^\mu_\perp,\, \delta( \omega + i \ndotD) \right\} \, \Gamma h_v, \nn\\
m_b \,  Q_2^{\mu}(\omega,\Gamma) &=&
 \bar{h}_v \left[i D^\mu_\perp,\delta(\omega + i \ndotD)\right] \, \Gamma \, h_v ,\\
m_b \,  Q_3(\omega,\Gamma) &=&
\!\!\int \!d \omega_1 d \omega_2 \, \delta(\omega_1, \omega_2; \omega)
 \bar{h}_v \, \delta(\omega_2 + i  \ndotD) g^{\mu\nu}_\perp \{ i D_\perp^\mu, i D_\perp^\nu \} 
\delta( \omega_1 + i \ndotD) \, \Gamma \, h_v, \nn\\
m_b \,  Q_4(\omega,\Gamma) &=&
\!\!-\!\!\int \!d \omega_1d \omega_2\,\delta(\omega_1, \omega_2; \omega)
 \bar{h}_v \, \delta( \omega_2 + i \ndotD)i\epsilon_\perp^{\mu\nu} [ i D_\perp^\mu, i D_\perp^\nu ] 
 \delta(\omega_1+ i \ndotD) \, \Gamma \, h_v ,\nn
\end{eqnarray}
where 
\begin{equation}
\delta(\omega_1, \omega_2; \omega) = \frac{\delta(\omega-\omega_1)- \delta(\omega-\omega_2)}{\omega_1 - \omega_2}.
\end{equation}
We define these operators rescaled by $m_b$ for later convenience in the anomalous dimension. This rescaling should be noted when comparing to other work dealing with subleading shape functions.
We also use the convention of labeling operators as $Q_i$ operators when the Dirac structure is general, and referring to the operators as $O_i$ and $P_i$ when a particular Dirac structure is required.  

We find that the operator basis must be extended beyond tree level to include,
at least the following operator 
\begin{eqnarray}
m_b \,  \bar{Q}_1^{\mu}(\omega,\Gamma) &=&
-2\!\! \int  \!d \omega_1 d \omega_2 \, \theta(\omega_1, \omega_2; \omega)
K_2^{\mu }(\omega_1, \omega_2; \Gamma),
\end{eqnarray}
where we have defined the following kernel and coefficient functions
\begin{equation}
K_2^{\mu}(\omega_1, \omega_2; \Gamma) =  \bar{h}_v \delta(\omega_1 + i \ndotD) i D_\perp^\mu \delta( \omega_2 + i \ndotD)
\Gamma h_v, 
\end{equation}
\begin{equation}
\theta(\omega_1, \omega_2; \omega) = \frac{\theta(\omega-\omega_1)- \theta(\omega-\omega_2)}{\omega_1 - \omega_2}.
\end{equation}

The operator $\bar{Q}_1$ was originally defined in \cite{blm01} by Mannel and Tackmann \cite{MannelTackman,Tackman05} based on symmetry arguments and examining the 
endpoint of $\btoc$ and taking the massless limit.  We find that the operator is unambiguously required beyond tree level due to the mixing experienced with the original set of operators. 

\subsection{Operator Feynman rules}

We use Feynman Gauge to calculate the 
anomalous dimension to subleading order as the 
usual choice of lightcone gauge introduces non physical poles in the calculation, for a review of the relevant issues see \cite{leibbrandt}.

Below, we present the required one and two gluon Feynman rules.
The two gluon Feynman rules are also required and used but are too lengthy to include here,
they can be obtained from the authors upon request.
The non-vanishing zero gluon Feynman rules in Feynman gauge are:
\begin{eqnarray} \label{zerogluonrules}
\langle h_v (k) | Q_0(\omega, \Gamma) | h_v(k) \rangle
&=& \delta(\omega + \ndotk) \Gamma , \nn \\
\langle h_v (k) | Q_1^\mu(\omega, \Gamma) ) | h_v(k) \rangle
&=& 2 \frac{k^\mu_\perp}{m_b} \delta(\omega + \ndotk) \Gamma , \nn \\
\langle h_v (k) | \bar{Q}_1^\mu(\omega, \Gamma) ) | h_v(k) \rangle
&=& 2  \frac{k^\mu_\perp}{m_b} \delta(\omega + \ndotk) \Gamma , \nn \\
\langle h_v (k) | Q_3(\omega, \Gamma) ) | h_v(k) \rangle
&=& -2  \frac{k_\perp^2}{m_b} \delta'(\omega + \ndotk) \Gamma.
\end{eqnarray}

The one gluon Feynman rule for the leading order operator is
\begin{eqnarray} \label{onegluonrulesshape}
\langle h_v (k) \, A^\nu_a(\ell)| Q_0(\omega, \Gamma) |h_v(k) \rangle  = - g T_a n^{\nu} 
	\left( \frac{\delta_-(\ndotl)}{\ndotl} \right) \Gamma . 
\end{eqnarray}
The one gluon Feynman rules for single covariant derivative operators are:
\begin{eqnarray}
\langle h_v (k) \, A^\nu_a(\ell)|   Q^{\mu}_1(\omega, \Gamma) | h_v(k) \rangle &=& - g T_a g^{\mu \nu}_{\perp}
	\delta_+(\ndotl) \frac{\Gamma}{m_b}  - g T_a n^{\nu} (2 k + \ell)^\mu_\perp
        \left( \frac{\delta_-(\ndotl)}{\ndotl} \right) \frac{\Gamma}{m_b} , \nn \\
 \langle h_v (k) \, A^\nu_a(\ell)|  \bar{Q}_1^{\mu}(\omega, \Gamma)  | h_v(k) \rangle &=& - 2g T_a g^{\mu \nu}_{\perp}
	\left( \frac{\theta_-(\ndotl)}{\ndotl} \right) \frac{\Gamma}{m_b} 
        + 2g T_a n^{\nu} \ell^\mu_\perp \left( \frac{\theta_-(\ndotl)}{(\ndotl)^2 } \right) \frac{\Gamma}{m_b} \nn \\
        &\,& \hspace{-2cm} -2g T_a n^{\nu} k^\mu_\perp \left( \frac{\delta_-(\ndotl)}{\ndotl} \right)\frac{\Gamma}{m_b} 
	-2g T_a n^{\nu} \ell^\mu_\perp \left( \frac{\delta(\omega + \ndotk + \ndotl)}{\ndotl} \right)\frac{\Gamma}{m_b} , \nn \\
\langle  h_v (k) \, A^\nu_a(\ell)| Q_2^{\mu}(\omega, \Gamma)  |  h_v(k) \rangle &=& g T_a g_{\perp}^{\mu \nu}
	\delta_-(\ndotl) \frac{\Gamma}{m_b} - g T_a n^{\nu} \ell^\mu_\perp
        \left( \frac{\delta_-(\ndotl)}{\ndotl} \right) \frac{\Gamma}{m_b} . 
\end{eqnarray}
Finally, the one gluon Feynman rules for two  covariant derivative operators are as follows:
\begin{eqnarray}
 \langle h_v (k) \, A^\nu_a(\ell)|   Q_3(\omega, \Gamma) | h_v(k) \rangle &=& 2 g T_a
 \left( (2 k + \ell)^{\nu}_\perp
	\left( \frac{\delta_-(\ndotl)}{\ndotl} \right)   - n^{\nu} k^2_\perp
        \left( \frac{\delta'(\omega + \ndotk)}{\ndotl} \right) \right) \frac{\Gamma}{m_b} \nn \\
        &\,&\hspace{-3.75cm} +2 g T_a
 \left( n^{\nu} (k+\ell)^2_\perp
        \left( \frac{\delta'(\omega + \ndotk + \ndotl)}{\ndotl} \right)
        -  n^{\nu} (2k_\perp \cdot \ell_\perp + \ell^2_\perp)
        \left( \frac{\delta_-(\ndotl)}{(\ndotl)^2} \right) \right) \frac{\Gamma}{m_b}, \nn \\
\langle h_v (k) \, A^\nu_a(\ell)|  Q_4(\omega, \Gamma)  | h_v(k)  \rangle &=& 2 g T_a i\epsilon_\perp^{\nu \beta} \ell_{\perp \beta}
\left( \frac{\delta_-(\ndotl)}{\ndotl} \right) \frac{\Gamma}{m_b} .
\end{eqnarray}
where $\ell$ is the gluon momentum flowing out of the vertex, and
the gluon carries Lorentz index $\nu$ and colour index $a$.
We have also made the convenient definitions
\begin{eqnarray} \label{deltapm}
\theta_\pm(x) &=& \theta(\omega+\ndotk+x) \pm \theta(\omega+\ndotk) \\
\delta_\pm(x) &=& \delta(\omega+\ndotk+x) \pm \delta(\omega+\ndotk). 
\end{eqnarray} 

\subsection{The Anomalous Dimension Matrix}

The renormalization of the operators $Q_i(\omega, \Gamma)$
is performed in the usual fashion,
\begin{equation}
\label{defZ}
Q_i(\omega, \Gamma)_\mathrm{bare}
=\int d\omega' Z_{ij}(\omega', \omega, \tilde\mu)
Q_i(\omega',  \tilde\mu, \Gamma)_\mathrm{ren},
\end{equation}
where $Z_{ij}(\omega', \omega, \tilde\mu)$ is a matrix of renormalization constants.
The values of the elements of $Z_{ij}$ can be found by
taking arbitrary partonic matrix elements of both sides, which at leading order gives
 $Z_{ij}^{(0)}(\omega', \omega, \tilde\mu)=
\delta_{ij}\delta(\omega-\omega')$.

To subleading order in $ \alpha_s $ we have
\begin{eqnarray}
\langle Q_i(\omega, \Gamma)\rangle_\mathrm{bare}^{(0)}+
\alpha_s \langle Q_i(\omega, \Gamma)\rangle_\mathrm{bare}^{(1)}
&=&\int d\omega' \left[Z_{ij}^{(0)}(\omega', \omega, \tilde\mu)+
\alpha_s Z_{ij}^{(1)}(\omega', \omega, \tilde\mu)\right]  \\
&&\hspace{1cm}\times \left[\langle Q_j(\omega', \tilde\mu, \Gamma)\rangle_\mathrm{ren}^{(0)}+
\alpha_s \langle Q_j(\omega', \tilde\mu, \Gamma)\rangle_\mathrm{ren}^{(1)}\right], \nn
\end{eqnarray}
from which one obtains
\begin{eqnarray}
\label{matchingequation}
\int d\omega' Z_{ij}^{(1)}(\omega', \omega, \tilde\mu)
\langle Q_j(\omega',  \tilde\mu, \Gamma)\rangle^{(0)}
&=& \langle Q_i(\omega, \Gamma)\rangle_\mathrm{bare}^{(1)}-
\langle Q_i(\omega, \Gamma)\rangle_\mathrm{ren}^{(1)} \nn \\
&=& (\langle Q_i(\omega, \Gamma)\rangle_\mathrm{bare}^{(1)})_{ \mathrm{div}}
\end{eqnarray}
where by $(\langle Q_i(\omega, \Gamma)\rangle_\mathrm{bare}^{(1)})_{ \mathrm{div}}$, we refer to the
UV divergent part of $\langle Q_i(\omega, \Gamma)\rangle_\mathrm{bare}^{(1)}$.
Because there are operators such as $Q_2$ and $Q_4$ which do not have a zero gluon form, we must consider matrix elements of Eq. (\ref{defZ}) with at least one external gluon.  These will be sufficient to identify the mixing of the various operators into $Q_2$ and $Q_4$.  It should be noted that  matrix elements with zero and one external gluon states are not sufficient in principle to determine the anomalous dimension matrix to subleading order. The operator
\begin{equation}
\label{bad_op}
Q^{\mu, \nu}(\omega_1, \omega_2, \Gamma) =
 \bar{h}_v \left[i D^\mu_\perp,\delta(\omega_2 + i \ndotD)\right] \left[i D^\nu_\perp,\delta(\omega_1 + i \ndotD)\right] \, \Gamma \, h_v ,
\end{equation}
does not have a zero gluon or one gluon Feynman rule. Its nonvanishing Feynman rules start at two gluon external states.  In this paper, we will not be calculating the two external gluon diagrams necessary to find mixing into this operator, if any exists.
We extract the anomalous dimension matrix of the subleading operators by examining 
matrix elements containing one perpendicularly polarized external gluon: 
\begin{equation}
\int d\omega' Z_{ij}^{(1)}(\omega', \omega, \tilde\mu)
\langle h_v(k)  A_{\perp}
 | Q_i(\omega', \Gamma)| h_v(k) \rangle^{(0)}_{\mathrm{ren}} = (\langle h_v(k) A_{\perp} |
 Q_i(\omega, \Gamma) | h_v(k) \rangle^{(1)}_{\mathrm{bare}})_{ \mathrm{div}}.
\end{equation}
The non-perpendicular components of the
gluon field were also examined but found to induce no further mixing.

The mixing of $Q_0$ into the other operators is determined by  calculating matrix elements 
of this operator with insertions of the subleading HQET Lagrangian. 
Zero gluon matrix elements are sufficient to find the mixing 
into $Q_0$, while one gluon matrix elements are required for mixing into
the remaining operators.  Due to the spin symmetry violating effects of 
the subleading HQET Lagrangian, the anomalous dimension of the $P_i$ operators can differ 
from that of the $O_i$ operators.

The wavefunction renormalization of the bare operators expressed
in terms of renormalized fields are
$Q_i(\omega, \Gamma)_\mathrm{bare}=Z_h Z_3^{n/2} Q_i(\omega, \Gamma)$ where
$n$ is the number of gluons in the operator, and $Q_i(\omega, \Gamma)$ is
written in terms of renormalized fields. For diagrams with an external state gluon 
we use the background field method to treat the external gluon as a classical
field and so we aquire no $Z_3$ factor due to the wavefunction renormalization of the gluon. \cite{abbott}


\subsection{Diagram Calculations}
\label{DiagramCalculations}


\subsubsection*{One Gluon Matrix Elements}
\begin{figure}[htbp]
\centerline{\scalebox{1.0}{\includegraphics{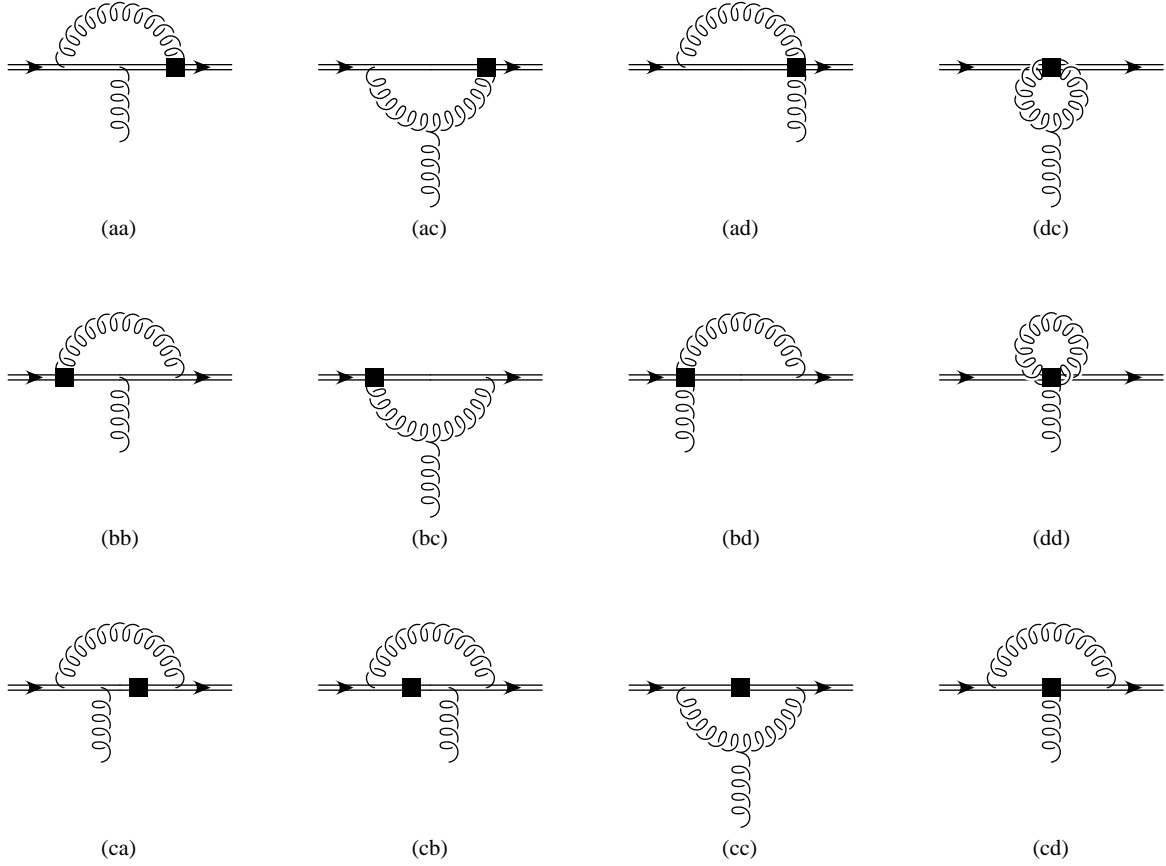}}}
\caption{The one gluon diagrams which must be calculated for each operator.}
\label{fig:gluondiag}
\end{figure}

The one gluon matrix elements are determined by calculating the
diagrams shown in Figure \ref{fig:gluondiag}
for each operator. The external gluon in each of these
diagrams is a background field gluon and the external states are chosen to 
have perpendicular polarization. We utilize dimensional regularization and the $\overline{\rm MS}$ scheme
to regulate our divergences.
To isolate and remove the IR divergences in the calculation
we keep {\it all} the particles off shell by retaining factors of $v\!\cdot\!k$,
$v\!\cdot\!\ell$ and $\ell^2$, where $\ell$ is the external gluon momentum.

To clearly illustrate the need to extend the operator basis we present the results for 
$Q_1$ diagram by diagram. In general, for perpendicular polarized external gluons, only diagrams (ac), (ad), (bc), (bd) and (dc) contribute to the amplitude. 
For diagram dc, the loop integrals to perform are as follows, with $c_1 \equiv \left(   \alpha_s \, g_s \, T_a \, g_{\perp}^{\mu \nu}\right) /(4 \, \pi )$,
\begin{eqnarray}
\langle  i\mathcal{A_{\mathrm{dc}}}  \rangle^{(1)} &=&- {\frac{c_1}{m_b} \, C_A \,(\ndotl)^2 \,  \mu^{4-d}}
\int\! \frac{d^d q}{(2\pi)^d} \frac{\delta_+(\ndotl)}
{\ndotq \, ( \ndotl + \ndotq ) \, (q^2 + i \epsilon )((q + \ell)^2 + i \epsilon)}  \\
&\,& + {\frac{c_1}{m_b} \, C_A \,(\ndotl)^2 \, \mu^{4-d}} \int\! \frac{d^d q}{(2\pi)^d} \frac{ \delta(\omega + \ndotk  - \ndotq ) +
 \delta(\omega + \ndotk +\ndotl + \ndotq ) }
{\ndotq \, ( \ndotl + \ndotq ) \, (q^2 + i \epsilon )((q + \ell)^2 + i \epsilon)}. \nn 
\end{eqnarray}
The integrals are perfomed via the standard techniques of dimensional regularization with $d = 4 - 2 \epsilon$, the $\rm \overline{MS}$ renormalization scale $\tilde\mu^2 = 4 \, \pi \mu^2 e^{ - \gamma_E} $, and the utilization of Eq.(\ref{thetaeqn}) and we have suppressed the lorentz and colour indicies.
The UV poles obtained for this diagram for $Q_1$ after 
the integrals are performed and consideration of the symmetry factor are
\begin{eqnarray} 
{\langle i \mathcal{A_{\mathrm{dc}}}  \rangle}_{ \mathrm{div}}^{(1)} &=& 
\frac{c_1 \, C_A}{ m_b \, \epsilon} \left(
\frac{ \ndotl }{\tilde\mu \,  ( \omega + \ndotk )} \phi_0'(\frac{\omega + \ndotk + \ndotl}{\tilde\mu})
- \frac{ \ndotl }{\tilde\mu \, ( \omega + \ndotk + \ndotl)} \phi_0'(\frac{\omega + \ndotk }{\tilde\mu})  \right) \nn \\
&\,& + \frac{c_1 \, C_A}{ m_b \, \epsilon} \left( \delta_+(\ndotl) \, \log(\frac{\ndotl}{\tilde\mu}) \right). \nn
\end{eqnarray}
The results for diagrams ac and bc when inserting $Q_1$ are
\begin{eqnarray}
{\langle i \mathcal{A_{\mathrm{ac}}} \rangle}_{ \mathrm{div}}^{(1)}
&=& \frac{c_1 \, C_A}{m_b \, \epsilon}
\left( \frac{\delta(\omega + \ndotk + \ndotl)}{2 \,\epsilon} +  \delta(\omega + \ndotk + \ndotl) \right) \nn \\
&\,& - \frac{c_1 \, C_A}{m_b \, \epsilon}  \left( \log(\frac{\ndotl}{\tilde\mu}) \, \delta(\omega + \ndotk + \ndotl) \right) \nn \\
&\,& + \frac{c_1 \, C_A}{m_b \, \ndotl \, \epsilon}
\left(\frac{(\omega + \ndotk)}{ (\omega + \ndotk + \ndotl)} \theta(\frac{\omega + \ndotk}{\tilde\mu})  
- \theta(\frac{\omega + \ndotk + \ndotl}{\tilde\mu}) \right), \\
{\langle i\mathcal{A_{\mathrm{bc}}} \rangle}_{ \mathrm{div}}^{(1)} &=&
\frac{ c_1 \, C_A}{m_b \, \epsilon}
\left( \frac{\delta(\omega + \ndotk)}{2 \,\epsilon} + \delta(\omega + \ndotk)
- \log(\frac{\ndotl}{\tilde\mu}) \, \delta(\omega + \ndotk) \right) \nn \\
&+&  \frac{ c_1 \, C_A}{ m_b \, \ndotl \, \epsilon}
\left(-\frac{(\omega + \ndotk + \ndotl)}{ (\omega + \ndotk)} \theta(\frac{\omega + \ndotk + \ndotl}{\tilde\mu}) 
+ \theta(\frac{\omega + \ndotk}{\tilde\mu}) \right).
\end{eqnarray}

Finally, the results for diagrams ad and bd for $Q_1$ insertions are
\begin{eqnarray}
{\langle  i\mathcal{A_{\mathrm{ad}}} \rangle}_{ \mathrm{div}}^{(1)} &=&
\frac{ c_1 \, C_F}{m_b \, \epsilon}
\left( \frac{\delta_{+}( \ndotl)}{ \epsilon}  - \frac{2}{\tilde\mu} \, \phi_0'(\frac{ \omega + \ndotk + \ndotl }{\tilde\mu})
 -  \frac{2}{\tilde\mu} \, \phi_0'(\frac{\omega + \ndotk}{\tilde\mu} )   \right) \nn \\
&+&  \frac{ c_1 \, C_A }{m_b \, \epsilon}
\left( - \frac{ \delta(\omega + \ndotk + \ndotl)}{2 \,\epsilon} +
 \frac{  \phi_0'((\omega + \ndotk + \ndotl)/ {\tilde\mu} ) }{\tilde\mu}  \right), \nn \\
{\langle i \mathcal{A_{\mathrm{bd}}}  \rangle}_{ \mathrm{div}}^{(1)}
&=& 
\frac{c_1 \, C_F}{m_b \, \epsilon} 
\left( \frac{\delta_{+}( \ndotl)}{ \epsilon}  -  \frac{2}{\tilde\mu} \, \phi_0'(\frac{\omega + \ndotk + \ndotl}{\tilde\mu} ) 
 -  \frac{2}{\tilde\mu} \, \phi_0'(\frac{ \omega + \ndotk}{\tilde\mu} )   \right) \nn \\
&+&  \frac{c_1 \, C_A }{m_b \, \epsilon} \left( - \frac{ \delta(\omega + \ndotk )}{2 \,\epsilon} + 
  \frac{  \phi_0'(( \omega + \ndotk)/{\tilde\mu} ) }{\tilde\mu}  \right).
\end{eqnarray}

The amplitudes combine to give the following UV poles
\begin{eqnarray}
{\langle  i \mathcal{A}_{Q_1} \rangle}_{ \mathrm{div}}^{(1)}  = 
\frac{c_1 \, C_F}{m_b \, \epsilon} 
\left( \frac{2 \, \delta_{+}( \ndotl)}{ \epsilon}  -  \frac{4 \, \phi'_{0+}( \ndotl /{\tilde\mu} ) }{\tilde\mu}  \right)
 + \frac{c_1  C_A }{ m_b \, \epsilon}  \left(  \delta_{+}( \ndotl) 
 + \frac{2 \, \theta_{-}(\ndotl /{\tilde\mu})}{\ndotl}  \right). 
\end{eqnarray}
Once the wavefunction renormalization terms are multiplicatively combined with the result, we 
express the amplitude in terms of renormalization matrix elements $d_i$ and the one gluon Feynman 
rules for the operators ${Q}_1^{\mu}$ and $\bar{Q}_1^{\mu}$ as follows
\begin{eqnarray}
{\langle  i \mathcal{A}_{Q_1} \rangle}_{ \mathrm{div}}^{(1)}
&=& \frac{\alpha}{4 \, \pi} 
\int  d \omega'  d_1 (\omega, \omega',\tilde\mu)  {\langle  Q_1^\mu(\omega') \rangle}^{(0)} \\
&\,& \! \! \! \!  + \frac{\alpha}{4 \, \pi} 
\int  d \omega' d_4 (\omega, \omega')  \left( {\langle  \bar{Q}_1^{\mu}(\omega') \rangle}^{(0)} - 
 {\langle  Q_1^\mu(\omega') \rangle}^{(0)} \right) , \nn
\end{eqnarray}
\noindent{where}
\begin{eqnarray}
d_1 (\omega, \omega',\tilde\mu) &=&  -\frac{2 C_F}{\epsilon^2}\delta(\omega-\omega')
+\frac{2 C_F}{\epsilon}\delta(\omega-\omega')
+\frac{4 C_F}{\tilde\mu \, \epsilon}
\phi_0'\left(\frac{\omega-\omega'}{\tilde\mu}\right), \nonumber \\
d_4 (\omega, \omega') &=& \frac{C_A}{\epsilon}\delta(\omega-\omega').
\end{eqnarray}
The form of the mixing between ${Q}_1^{\mu}$ and $\bar{Q}_1^{\mu}$ deserves some comment.  
At zero gluon the matrix elements of these operators are identical causing this mixing to be undetermined for zero gluon external state diagrams, even though the zero 
gluon matrix elements of both operators are nonzero, contrary to naive expectations.
The contribution of the operator  $\bar{Q}_1^{\mu}$ to the renormalization matrix was also determined.
We find that this operator mixes with itself contributing a $d_1$ form to the matrix $Z_{SL}$.  
The antisymmetric operators $ Q_2$ and $Q_4$ mix only with themselves and 
contribute diagonal factors of $d_1$ to the matrix of renormalization constants. 

The inauspicious form of mixing between $Q_1^{\mu}$ and $\bar{Q}_1^{\mu}$ is also present 
in the operator $Q_3$, however the corresponding analysis of  $Q_3$ is more complicated and is still under investigation. Due to this complication in determining the full anomalous dimension matrix and the need for a two gluon calculation to determine the possible mixing with $ Q^{\mu, \nu}(\omega_1, \omega_2, \Gamma) $ we present the results of our initial study of the anomalous dimension to subleading order in this paper and comment on the phenomenological consequences of the presented results. 
We collect our results in section \ref{results}.


\subsubsection*{The T products of ${\cal{L}}_{HQET}$  with $O_0$ and $P_0$ }
To find the mixing of the operators
\begin{eqnarray}
O_0 &=& \bar{h}_v \, \delta(\omega + i \ndotD ) \, h_v, \nonumber \\
P_0^{\mu} &=& \bar{h}_v \, \delta(\omega + i \ndotD ) \, \gamma^{\mu} \gamma_5 \, h_v 
\end{eqnarray}
into the subleading operators, we must evaluate the time ordered products of the operators with the
the subleading terms of the HQET Lagrangian ($\mathcal{L}_1$)
\begin{eqnarray}
T_O(\omega) &=& i \,  \int d^4 x \, T \left[ \mathcal{L}_1(x), O_0(0,\omega) \right], \nn \\
T_P^{\mu}(\omega) &=& i \, \int d^4 x \, T \left[ \mathcal{L}_1(x), P_0^{\mu}(0,\omega) \right]. 
\end{eqnarray}
We now explicitly refer to the Dirac structure of the operators.  This is necessary due to the Dirac structure of the operators in the subleading Lagrangian. We treat the subleading Lagrangian as a single operator insertion for the purposes of our calculation. 
The different renormalization of the kinetic and chromomagnetic terms is accommodated
by breaking the T products up in to $ T_{(O_0,O_k)},T_{(O_0,O_m)},T_{(P_0,O_k)},T_{(P_0,O_m)}$ after the diagram calculations, where $O_k, O_m$ refer to the kinetic and chromomagnetic operators of the subleading Lagrangian. 

\begin{figure}[htbp]
\centerline{\scalebox{1.0}{\includegraphics{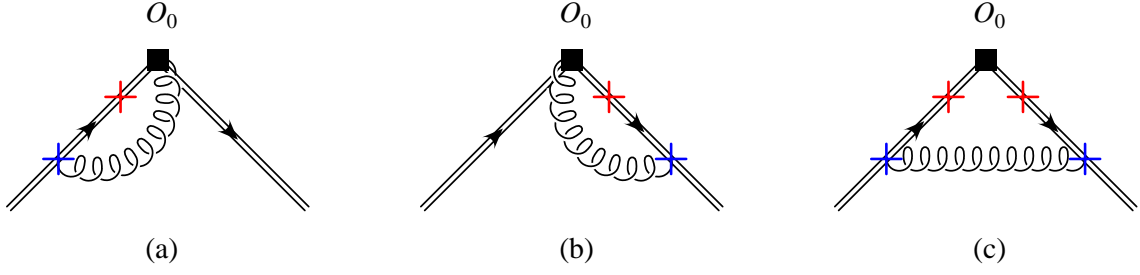}}}
\caption{The diagrams which must be calculated for $Q_0$.  The
crosses represent insertions of the subleading HQET Lagrangian.}
\label{fig:threediags}
\end{figure}

We start with the zero gluon diagrams.  They are
illustrated in Figure \ref{fig:threediags}.
The crosses in the diagrams represent the locations where one inserts 
the subleading HQET Lagrangian, given by
\begin{equation}
\mathcal{L}_1 = \bar{h}_v \frac{(i \, D_\perp)^2}{2 m_b} h_v - a(\tilde{\mu})
 \bar{h}_v \frac{(g \, \sigma_{\alpha \, \beta} \, G^{\alpha \, \beta})}{4 m_b} h_v.
\end{equation}
The zero, one and two gluon Feynman rules for this Lagrangian,
where we suppress the renormalization scale dependence 
of the $O_m$ operator, are
\begin{eqnarray}
i\mathcal{L}_1[\mbox{0-gluon}]&=&i\frac{k_\perp^2}{2m_b}P_+\nonumber\\
i\mathcal{L}_1[\mbox{1-gluon}]&=&
-i g T_a \frac{(2k_\perp+\ell_\perp)^\alpha}{2m_b} P_+
+i g T_a \frac{i\epsilon^{\alpha \mu \rho \eta}
l_\mu v_\rho}{2m_b} s_\eta\nonumber\\
i\mathcal{L}_1[\mbox{2-gluon}]&=&
i g^2 \{T^{a_1}, T^{a_2}\} \frac{g_{\perp}^{\alpha_1 \alpha_2}}{2m_b}P_+
+i g^2 [T^{a_1}, T^{a_2}]
\frac{i\epsilon^{\alpha_1 \alpha_2 \rho \eta}v_\rho}{2m_b} s_\eta.
\end{eqnarray}

The UV divergent part of the sum of the subleading ${\cal{L}}_{\mathrm{HQET}}$ zero gluon results 
for both $O_0$ and $P_0$ (up to Dirac structure)  is the same. 
Our result for the general Dirac structure operator $Q_0$, with indices suppressed, is
\begin{eqnarray}
\langle h_v(k) | T_{(Q_0,O_k)}(\omega,\Gamma) | h_v(k)  \rangle_\mathrm{div}^{(1)}
& = & \frac{\alpha_s}{4 \, m_b \, \pi} \, C_F
\int d\omega'  \frac{4\,\omega' \delta(\omega-\omega')}{\epsilon}
\langle h_v(k) | Q_0(\omega', \Gamma)  | h_v(k) \rangle^{(0)}  \nonumber \\
&\,& \! \! \! \! \! \! \! \! \! \! \! \! + \frac{\alpha_s}{4\pi} \, C_F
\int d\omega'   \frac{3\, \delta(\omega-\omega')}{\epsilon} v_{\mu}
\langle h_v(k) |  Q_1^{\mu}(\omega', \Gamma)  | h_v(k) \rangle^{(0)}.
\end{eqnarray}
The latter term 
is the zero gluon matrix element of a modified $Q_1^{\mu}$ operator. 
The operator is modified to not have a perpendicular covariant derivative, 
but simply a $D^{\mu}$ in its definition. This term vanishes at this order due 
to the equations of motion, these results are combined with our operator anomalous dimensions in section \ref{results}. (It should be noted that the zero gluon calculation does not determine this mixing is with the $Q_1$ operator, as its zero gluon rule is identical to $\bar{Q}_1$. Consistency between zero and one gluon external state calculations determines this mixing to be with the modified $Q_1$ operator.)

\begin{figure}[htbp]
\centerline{\scalebox{1.0}{\includegraphics{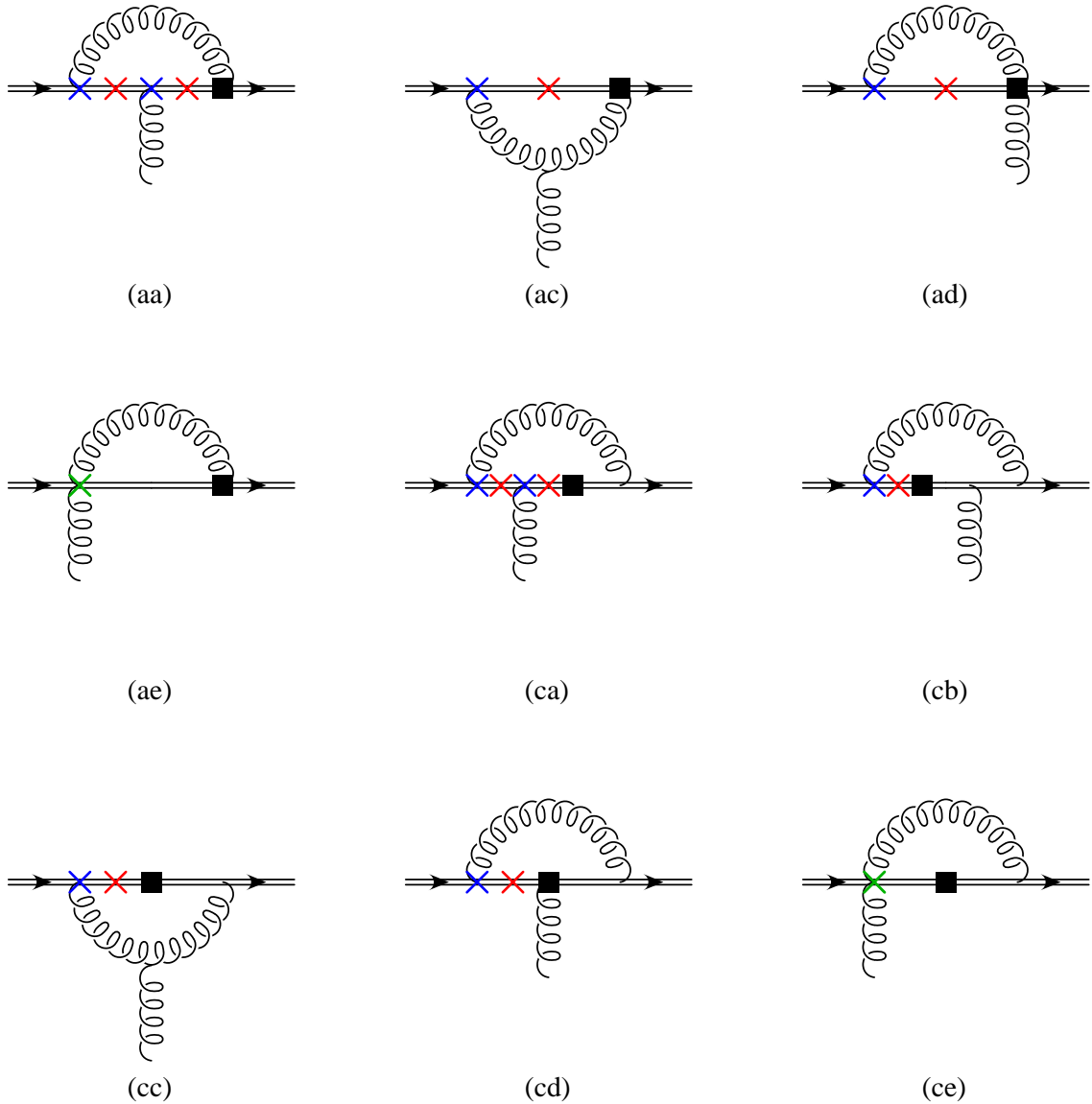}}}
\caption{The one gluon diagrams which must be calculated for $O_0$
and $P_0$.  The crosses represent insertions of the subleading HQET Lagrangian.
The mirror diagram corresponding to each of the above diagrams is not shown.}
\label{fig:ninediags}
\end{figure}

There are many more one gluon diagrams than zero gluon diagrams,
as illustrated in Figure \ref{fig:ninediags}.
The diagrams explicitly given constitute half of the
total number of diagrams that must be calculated for each of $O_0$
and $P_0$.  The other diagrams can be looked upon as either the
mirror diagrams of those given, or the transposed diagrams which have
the $\mathcal{L}_1$ operator and lightcone operator interchanged. 

The Dirac structure of the subleading
Lagrangian force us to consider the Dirac structure of these diagrams.
Let us consider the one gluon diagrams of Figure \ref{fig:ninediags},
where the lightcone operator is $Q_0$.  We will denote by
$  \langle ..| \mathcal{A}_R  |..\rangle $ and $  \langle ..| \mathcal{A}_L  | .. \rangle $ the
amplitude of these diagrams and the amplitude of the mirror diagrams respectively. 
Because of the simple relationship between $O_0$ and
$P_0^\sigma$, the corresponding amplitues for $P_0^\sigma$ are
$ \langle .. |s^\sigma \mathcal{A}_R  | .. \rangle $ and 
$\langle .. |  \mathcal{A}_L s^\sigma | .. \rangle $.

Each of $\langle \mathcal{A}_R \rangle$ and $ \langle \mathcal{A}_L \rangle$ can be decomposed
into the two Dirac structures $P_+$ and $s^\eta$, for example with a heavy quark target, with one gluon in the final external state:
\begin{eqnarray}
\langle   h_v(k) \, A^\nu_a | \mathcal{A}_R  | h_v(k) \rangle &=& \langle   h_v(k) \, A^\nu_a | \mathcal{A}_{R+} P_+  | h_v(k) \rangle
+ \langle   h_v(k) \, A^\nu_a | \mathcal{A}_{Rs}^{ \eta} s_\eta  | h_v(k) \rangle, \nonumber \\
\langle   h_v(k) \, A^\nu_a | \mathcal{A}_L  | h_v(k) \rangle &=& \langle   h_v(k) \, A^\nu_a | \mathcal{A}_{L+} P_+  | h_v(k) \rangle
+ \langle   h_v(k) \, A^\nu_a | \mathcal{A}_{Ls}^{\eta} s_\eta  | h_v(k) \rangle.
\end{eqnarray}

Thus for operator $O_0$ we can write the total amplitude proportional to
each of the Dirac structures after the calculations of the 40 diagrams
required. The results for insertions of the $O_k$  are:
\begin{eqnarray}
 \langle   h_v(k) \, A^\nu_a | T_{(O_0,O_k)} | h_v(k) \rangle_{\mathrm{div}}^{(1)} 
 &=&  \langle   h_v(k) \, A^\nu_a |  (\mathcal{A}_{R+} +
\mathcal{A}_{L+}) P_+ | h_v(k)  \rangle_{\mathrm{div}}^{(1)}, \nonumber \\
&=&-  \frac{C_F \, \omega \, \alpha_s g_s}{\pi \, m_b \, \epsilon }\,n^{\nu} T_a 
\left(\frac{\delta _{-} \left( \ndotl \right)}{\ndotl}\right) P_+ \\  
&\,& \hspace{-1cm} -  \frac{3 \, C_F \, \alpha_s g_s}{4 \, \pi \, m_b \, \epsilon } \, T_a \,
\left(v^{\nu} \delta_{+} \left( \ndotl \right)  - n^{\nu} 
\left( 2 \, k \cdot v + \ell \cdot v \right)\left( \frac{\delta_{+} \left( \ndotl \right)}{\ndotl}\right) 
\right) P_+. \nonumber 
\end{eqnarray}
This result is easily matched, it is identical to the mixing form found in the 
zero gluon result with $\Gamma = P_+$, so that the mixing occurs with the 
operators $O_0$ and $O_1$:
\begin{eqnarray}
 \langle h_v(k) \, A^\nu_a | T_{(O_0,O_k)} | h_v(k) \rangle_{\mathrm{div}}^{(1)} 
& = & \frac{\alpha_s}{4 \, \pi} \, C_F
\int d\omega'  \frac{4\,\omega' \delta(\omega-\omega')}{\epsilon\, m_b}
\langle h_v(k) \, A^\nu_a |  O_0(\omega')  | h_v(k) \rangle^{(0)}  \nn\\
&\,&  \hspace{-1cm} +  \frac{\alpha_s}{4\pi} \, C_F
\int d\omega'  \frac{3\, \delta(\omega-\omega')}{\epsilon} v_{\mu}
\langle h_v(k)  \, A^\nu_a |  O_1^{\mu}(\omega')  | h_v(k) \rangle^{(0)}.
\end{eqnarray}
The result of the T product with $O_m$ is 
\begin{eqnarray}
 \langle h_v(k) \, A^\nu_a | T_{(O_0,O_m)} | h_v(k) \rangle_{\mathrm{div}}^{(1)}
&=& \langle h_v(k) \, A^\nu_a | [\mathcal{A}_{Rs}^{\eta} + 
\mathcal{A}_{Ls}^{ \eta}] s_\eta  | h_v(k) \rangle_{\mathrm{div}}^{(1)}, \nonumber \\
&=& - \frac{C_A \, g_s \, \alpha \, T^a}{8 \, m_b \,  \epsilon \, \pi } 
 \left( i \,\epsilon_\perp^{\nu s}  \delta_{-}\left( \ndotl \right) 
 - i \,\epsilon_\perp^{\ell \, s} \, n^{\nu} \,\frac{\delta_{-}\left( \ndotl \right)}{ \ndotl} \right).
\end{eqnarray}
This result matches onto the one gluon rule for $P_2$, as expected by the 
symmetry of the single derivative operators:
\begin{eqnarray}
\langle h_v(k) \, A^\nu_a |  T_{(O_0,O_m)}| h_v(k) \rangle_{\mathrm{div}}^{(1)}
& = & - \frac{\alpha_s}{4\pi} \, C_A
\int d\omega'   \frac{\delta(\omega-\omega') }{2\, \epsilon}  \, i \, \epsilon_\perp^{\mu \, \sigma} \, \langle h_v(k) \, A^\nu_a |  P_{2 \mu \, \sigma}( \omega' ) | h_v(k)  \rangle^{(0)}.\nn
\end{eqnarray}

The total $T_{(P_0,O_k)}^\sigma $ and $T_{(P_0,O_m)}^\sigma $ amplitudes can be written as
\begin{eqnarray}
 \langle h_v(k) \, A^\nu_a | T_{(P_0,O_k)}^\sigma  | h_v(k) \rangle &=& 
  \langle h_v(k) \, A^\nu_a |s^\sigma \mathcal{A}_{R+} | h_v(k) \rangle
+ \langle h_v(k) \, A^\nu_a | s^\sigma \mathcal{A}_{L+} | h_v(k) \rangle, \\
 \langle h_v(k) \, A^\nu_a | T_{(P_0,O_m)}^\sigma  | h_v(k) \rangle &=& 
 \langle h_v(k) \, A^\nu_a |s^\sigma s^\eta \mathcal ({A}_{Rs})_{\eta} | h_v(k) \rangle  + \langle h_v(k) \, A^\nu_a | s^\eta s^\sigma \mathcal ({A}_{Ls})_{\eta} | h_v(k) \rangle. \nn
\end{eqnarray} 
Using the decomposition
\begin{equation}
\label{ssDecomposition}
s^\sigma s^\rho = i \epsilon^{\sigma \rho \eta \phi} s_\eta v_\phi
-(g^{\sigma \rho}-v^\sigma v^\rho)P_+,
\end{equation}
we can decompose in terms of the pieces
proportional to $P_+$ and $s^\eta$ for these amplitudes. For  $T_{(P_0,O_k)}^\sigma$ and $T_{(P_0,O_m)}^\sigma$ we find the following mixing
\begin{eqnarray}
\langle h_v(k) \, A^\nu_a | T_{(P_0,O_k)}^\sigma | h_v(k)  \rangle_{\mathrm{div}}^{(1)}
&=& \langle  h_v(k) \, A^\nu_a | s^\sigma (\mathcal{A}_{R+} +\mathcal{A}_{L+}) | h_v(k)  \rangle, \nn \\
&=&  \frac{\alpha_s}{4 \, \pi} \, C_F
\int d\omega'  \frac{4\,\omega' \delta(\omega-\omega')}{\epsilon \, m_b}
\langle h_v(k) \, A^\nu_a |  P_0(\omega')^\sigma   | h_v(k) \rangle^{(0)}  \nn\\
&\,&  \! \!  \! \! \! +  \frac{\alpha_s}{4 \,\pi} \, C_F
\int d\omega'  \frac{3\, \delta(\omega-\omega')}{\epsilon} v_{\mu}
\langle h_v(k)  \, A^\nu_a |  P_1^{\mu}(\omega')^\sigma   | h_v(k) \rangle^{(0)}, \nn 
\end{eqnarray}
\begin{eqnarray}
\langle h_v(k) \, A^\nu_a |  T_{(P_0,O_m)}^\sigma  | h_v(k)  \rangle_{\mathrm{div}}^{(1)}
&=& (v^\sigma v^\mu - g^{\sigma \, \mu})
\langle h_v(k) \, A^\nu_a | (\mathcal ({A}_{Rs})_{\mu} + (\mathcal{A}_{Ls})_{\mu}) P_+ | h_v(k)  \rangle
\nn \\ & \,& + \langle  h_v(k) \, A^\nu_a | [
i \epsilon^{\sigma\eta\phi\mu} \, v_\mu \,
(\mathcal ({A}_{Rs})_{\eta}-\mathcal ({A}_{Ls})_{\eta})
 s_\phi | h_v(k) \rangle, \nn\\
&=& \frac{\alpha_s}{4 \, \pi} \, C_A
\int d\omega'   \frac{\delta(\omega-\omega') }{2\, \epsilon} \, i \, \epsilon_\perp^{\mu \, \sigma} \,
 \langle h_v(k) \, A^\nu_a |O_{2 \mu}( \omega' ) | h_v(k) \rangle  \nn \\
&\,&  \! \!  \! \! \! -  \frac{\alpha_s}{4 \, \pi}  \, C_A
\int d\omega'  \frac{ \delta(\omega-\omega')}{2 \, \epsilon} 
\left[ (v^\sigma  - n^\sigma) g^{\mu \, \eta} + n^{\eta} ( g^{\sigma \mu} - v^\sigma \, v^\mu) \right] \times \nn \\
&\,&  \! \!  (\langle h_v(k) \, A^\nu_a |  P_1^{\mu \, \eta} (\omega')  | h_v(k) \rangle^{(0)}  - 
\langle h_v(k) \, A^\nu_a |  \bar{P}_1^{\mu \, \eta} (\omega')  | h_v(k) \rangle^{(0)} ). \nn
\end{eqnarray}
\section{Results}
\label{results}
\subsection{Leading Nonperturbative Order}
The order $\alpha_s$ perturbative and leading order non-perturbative
anomalous dimension matrix has been calculated by a variety of  authors \cite{Bauer:2003pi,BoschBjornNeubertPaz0402}.
Our results agree with theirs, and in the basis
\begin{equation}
\{O_0, P_0\}
\end{equation}
the pertrubative expansion is given by
\begin{eqnarray}
{Z^{(0)}}(\omega, \omega') &=&   \left[
\begin{array}{cc}
\delta(\omega-\omega') & 0\\
0 & \delta(\omega-\omega') \\
\end{array}
\right],  \\
Z^{(1)}(\omega, \omega', \tilde\mu)&=& \frac{\alpha_s(\tilde\mu)}{4 \pi}\left[
\begin{array}{cc}
d_1(\omega, \omega', \tilde\mu) & 0 \\
0 & d_1(\omega, \omega', \tilde\mu) \\
\end{array}
\right].
\label{Zmatrix0}
\end{eqnarray}
The distribution $d_1(\omega, \omega',\tilde\mu) $ is the combination of the operator and wavefunction renormalization counter terms, given by
\begin{eqnarray}
d_1(\omega, \omega',\tilde\mu) &=& -\frac{2 C_F}{\epsilon^2}\delta(\omega-\omega')
+\frac{2 C_F}{\epsilon}\delta(\omega-\omega')
+\frac{4 C_F}{\tilde\mu\,\epsilon}
\phi_0'\left(\frac{\omega-\omega'}{\tilde\mu}\right).
\end{eqnarray}
Recall, our initial expression related the bare and renormalized operators,
\begin{equation}
Q_0(\omega, \Gamma)_\mathrm{bare}
=\int d\omega' Z(\omega', \omega,\tilde\mu)
Q_0(\omega', \tilde\mu, \Gamma)_\mathrm{ren}.
\end{equation}
We differentiate this equation with respect to $\tilde\mu$ to obtain our renormalization group equation
\begin{equation}
\tilde\mu \frac{d}{d \tilde\mu}  Q_0(\omega, \tilde\mu,  \Gamma)_\mathrm{ren}
= - \int d\omega' \gamma(\omega', \omega, \tilde\mu)
Q_0(\omega',\tilde\mu, \Gamma)_\mathrm{ren}.
\end{equation}
The anomalous
dimension matrix is determined using the useful result for $\overline{\rm MS}$ \cite{Floratos} 
\begin{eqnarray}\label{anomdimalpha}
\gamma(g_s) = - 2 \alpha_s \frac{d Z_1(\alpha_s)}{d \alpha_s},
\end{eqnarray}
where $Z_1$ is the coefficient of the $1/ \epsilon$ poles. We find
\begin{equation}
\gamma(\omega, \omega',\tilde\mu)=
\left[
\begin{array}{cc}
\gamma_1(\omega, \omega',\tilde\mu) & 0 \\
0 & \gamma_1(\omega, \omega',\tilde\mu) \\
\end{array}
\right],
\end{equation}
with,
\begin{equation}
\gamma_1(\omega, \omega',\tilde\mu)=- \frac{\alpha_s(\tilde\mu)}{\pi}C_F\left[
\delta(\omega-\omega')
+\frac{2}{\tilde\mu}
\phi_0'\left(\frac{\omega-\omega'}{\tilde\mu}\right)
\right].
\end{equation}
\subsection{Subleading Nonperturbative Order}
We have determined the  matrix of renormalization constants at subleading non-perturbative order ${Z_{SL}}$,  excluding operators of class $Q_3$ .
If we order our $Q_i$ operators as
\begin{eqnarray}
\mathcal{O}_i &=& \{O_0,T_{(O_0,O_k)},T_{(O_0,O_m)}, O_1^\mu, \bar{O}_1^\mu, O_2^\mu,  O_4\},\nonumber \\
\mathcal{P}_i^\sigma &=& \{P_0^\sigma,T_{(P_0,O_k)}^\sigma,T_{(P_0,O_m)}^\sigma, P_1^{\sigma \mu}, \bar{P}_1^{\sigma \mu}, P_2^{\sigma \mu},  P_4^\sigma \}, 
\end{eqnarray}
the leading order term in the perturbative expansion in the basis $\{\mathcal{O}_i, \mathcal{P}_i\}$ is
given in block form as
\begin{eqnarray}
{Z^{(0)}_{SL}}(\omega, \omega') =  \left( 
\begin{array}{cc}
\Gamma_{O_i,O_j}^0(\omega, \omega') & \Gamma_{O_i,P_j}^0(\omega, \omega') \\
\Gamma_{P_i,O_j}^0(\omega, \omega') & \Gamma_{P_i,P_j}^0(\omega, \omega') \\
\end{array}
\right). 
\end{eqnarray}
Where the entries in the matrices in the above expression with $i,j = 0,1...6$ are given by
\begin{eqnarray}
\Gamma_{O_i,O_j}^0(\omega, \omega') &=& (\delta_{i,j} - \delta_{0,j}\delta_{i,0}) \delta(\omega - \omega'), \nn \\
\Gamma_{P_i,P_j}^0(\omega, \omega') &=& (\delta_{i,j} - \delta_{0,j}\delta_{i,0}) \delta(\omega - \omega'), \nn \\
\Gamma_{O_i,P_j}^0(\omega, \omega') &=& 0,  \\
\Gamma_{P_i,O_j}^0(\omega, \omega') &=& 0.  \nn
\end{eqnarray}

While the $\orderalpha$ term in the expansion is
\begin{eqnarray}
Z^{(1)}_{SL}(\omega, \omega',\tilde\mu)= \frac{\alpha_s(\tilde\mu)}{4 \pi}  \left( 
\begin{array}{cc}
\Gamma_{O_i,O_j}^1(\omega, \omega') & \Gamma_{O_i,P_j}^1(\omega, \omega') \\
\Gamma_{P_i,O_j}^1(\omega, \omega') & \Gamma_{P_i,P_j}^1(\omega, \omega') \\
\end{array}
\right). 
\end{eqnarray}
with the diagonal block matrices 
\begin{eqnarray}
\Gamma_{O_i,O_j}^1(\omega, \omega') =  \left[
\begin{array}{ccccccc}
0 & 0 & 0 & 0 & 0 & 0 & 0  \\
d_2(\omega, \omega',\tilde\mu)  & 0  & 0 & d_3^{\mu}(\omega, \omega')   & 0 & 0 & 0  \\
0 & 0  & 0 & 0 & 0 & 0 & 0  \\
0 & 0
& 0 & d_4(\omega, \omega',\tilde\mu)   & d_5(\omega, \omega')  & 0 & 0 \\
0 & 0  & 0 & 0  &  d_1(\omega, \omega',\tilde\mu)  & 0 & 0  \\
0 & 0  & 0 & 0  & 0 &  d_1(\omega, \omega',\tilde\mu) & 0 \\
0 & 0  & 0 & 0  & 0 & 0 &  d_1(\omega, \omega',\tilde\mu)  
\end{array}
\right], \nn
\end{eqnarray}
\begin{eqnarray}
\Gamma_{P_i,P_j}^1(\omega, \omega') =  \left[
\begin{array}{ccccccc}
0 & 0 & 0 & 0 & 0 & 0 & 0  \\
d_2(\omega, \omega',\tilde\mu)  & 0  & 0 & d_3^{\mu}(\omega, \omega')   & 0 & 0 & 0  \\
0 & 0  & 0 & - d_7^{\mu \, \sigma \, \eta}(\omega, \omega') & d_7^{\mu \, \sigma \, \eta}(\omega, \omega') & 0 & 0  \\
0 & 0
& 0 & d_4(\omega, \omega',\tilde\mu)   & d_5(\omega, \omega')  & 0 & 0 \\
0 & 0  & 0 & 0  &  d_1(\omega, \omega',\tilde\mu)  & 0 & 0  \\
0 & 0  & 0 & 0  & 0 &  d_1(\omega, \omega',\tilde\mu) & 0 \\
0 & 0  & 0 & 0  & 0 & 0 &  d_1(\omega, \omega',\tilde\mu)  
\end{array}
\right]. \nn
\end{eqnarray}
The off diagonal block matrices are as follows
\begin{eqnarray}
\Gamma_{O_i,P_j}^1(\omega, \omega') =  \left[
\begin{array}{ccccccc}
0 & 0  & 0 & 0 & 0 & 0 & 0  \\
0 & 0  & 0 & 0 & 0 & 0 & 0  \\
0 & 0  & 0 & 0 & 0 &  -d_6^{\mu \, \sigma}(\omega, \omega')  & 0  \\
0 & 0  & 0 & 0 & 0 & 0 & 0  \\
0 & 0  & 0 & 0 & 0 & 0 & 0  \\
0 & 0  & 0 & 0 & 0 & 0 & 0  \\
0 & 0  & 0 & 0 & 0 & 0 & 0  
\end{array}
\right], \nn
\end{eqnarray}
\begin{eqnarray}
\Gamma_{P_i,O_j}^1(\omega, \omega') =  \left[
\begin{array}{ccccccc}
0 & 0  & 0 & 0 & 0 & 0 & 0  \\
0 & 0  & 0 & 0 & 0 & 0 & 0  \\
0 & 0  & 0 & 0 & 0  & d_6^{\mu \, \sigma}(\omega, \omega') & 0  \\
0 & 0  & 0 & 0 & 0 & 0 & 0  \\
0 & 0  & 0 & 0 & 0 & 0 & 0  \\
0 & 0  & 0 & 0 & 0 & 0 & 0  \\
0 & 0  & 0 & 0 & 0 & 0 & 0  
\end{array}
\right]. \nn
\end{eqnarray}
The  $d_i(\omega, \omega',\tilde\mu) $ distributions are given by
\begin{eqnarray}
d_2(\omega, \omega', \tilde\mu) &=&  \frac{4 \, C_F }{ \epsilon} \frac{\omega'}{m_b(\tilde\mu)}\delta(\omega-\omega'), \nonumber\\
d_3^{\mu}(\omega, \omega') &=&  \frac{3 \, C_F }{ \epsilon} \delta(\omega-\omega') v^{\mu}, \nonumber\\
d_4(\omega, \omega', \tilde\mu) &=& d_1(\omega, \omega',\tilde\mu)  - \frac{C_A}{ \epsilon}\delta(\omega-\omega'), \nonumber\\
d_5(\omega, \omega') &=& \frac{ C_A}{\epsilon}\delta(\omega-\omega'), \nonumber\\
d_6^{\mu \, \sigma}(\omega, \omega') &=&  -\frac{C_A }{2 \, \epsilon} ( i \, \epsilon^{\mu \, \sigma}_{\perp}) \delta(\omega-\omega'), \nonumber\\
d_7^{\mu \, \sigma \, \eta}(\omega, \omega') &=&  \frac{C_A }{2 \, \epsilon} ((v^\sigma  - n^\sigma) g^{\mu \, \eta} + n^{\eta} ( g^{\sigma \mu} - v^\sigma \, v^\mu)) \delta(\omega-\omega'). 
\end{eqnarray}
We directly determine the
anomalous dimension matrix to subleading order using Eq. (\ref{anomdimalpha}) to be the following
\begin{eqnarray}
\gamma_{SL}(\omega, \omega',\tilde\mu)= \left(
\begin{array}{cc} 
\gamma_{O_i,O_j}(\omega, \omega',\tilde \mu) & \gamma_{O_i,P_j}(\omega, \omega',\tilde \mu) \\
\gamma_{P_i,O_j}(\omega, \omega',\tilde \mu) & \gamma_{P_i,P_j}(\omega, \omega',\tilde \mu) \\
\end{array}
\right). 
\end{eqnarray}
The diagonal entries of the anomalous dimension matrix are
\begin{eqnarray}
\gamma_{O_i,O_j}(\omega, \omega',\tilde \mu) =  \left[
\begin{array}{ccccccc}
0 & 0 & 0 & 0 & 0 & 0 & 0  \\
 \gamma_2(\omega, \omega',\tilde\mu)  & 0  & 0 &  \gamma_3^{\mu}(\omega, \omega',\tilde \mu)  & 0 & 0 & 0  \\
0 & 0  & 0 & 0 & 0 & 0 & 0  \\
0 & 0 & 0 & \gamma_4(\omega, \omega',\tilde\mu)   & \gamma_5(\omega, \omega',\tilde \mu)  & 0 & 0 \\
0 & 0  & 0 & 0  &  \gamma_1(\omega, \omega',\tilde\mu)  & 0 & 0  \\
0 & 0  & 0 & 0  & 0 &  \gamma_1(\omega, \omega',\tilde\mu) & 0 \\
0 & 0  & 0 & 0  & 0 & 0 &  \gamma_1(\omega, \omega',\tilde\mu)  
\end{array}
\right], \nn
\end{eqnarray}
\begin{eqnarray}
\gamma_{P_i,P_j}(\omega, \omega',\tilde \mu) =  \left[
\begin{array}{ccccccc}
0 & 0  & 0 & 0 & 0 & 0 & 0  \\
 \gamma_2(\omega, \omega',\tilde\mu)   & 0  & 0 &  \gamma_3^{\mu}(\omega, \omega',\tilde \mu)   & 0 & 0 & 0  \\
0 & 0  & 0 & - \gamma_7^{\mu \, \sigma \, \eta}(\omega, \omega',\tilde \mu) & \gamma_7^{\mu \, \sigma \, \eta}(\omega, \omega',\tilde \mu) & 0 & 0  \\
0 & 0
& 0 & \gamma_4(\omega, \omega',\tilde\mu)   & \gamma_5(\omega, \omega',\tilde \mu)  & 0 & 0 \\
0 & 0  & 0 & 0  &  \gamma_1(\omega, \omega',\tilde\mu)  & 0 & 0  \\
0 & 0  & 0 & 0  & 0 &  \gamma_1(\omega, \omega',\tilde\mu) & 0 \\
0 & 0  & 0 & 0  & 0 & 0 &  \gamma_1(\omega, \omega',\tilde\mu)  
\end{array}
\right]. \nn
\end{eqnarray}
The off diagonal entries are as follows
\begin{eqnarray}
\gamma_{O_i,P_j}(\omega, \omega',\tilde \mu) =  \left[
\begin{array}{ccccccc}
0 & 0  & 0 & 0 & 0 & 0 & 0  \\
0 & 0  & 0 & 0 & 0 & 0 & 0  \\
0 & 0  & 0 & 0 & 0 &  -\gamma_6^{\mu \, \sigma}(\omega, \omega')  & 0  \\
0 & 0  & 0 & 0 & 0 & 0 & 0  \\
0 & 0  & 0 & 0 & 0 & 0 & 0  \\
0 & 0  & 0 & 0 & 0 & 0 & 0  \\
0 & 0  & 0 & 0 & 0 & 0 & 0  
\end{array}
\right], \nn
\end{eqnarray}
\begin{eqnarray}
\gamma_{P_i,O_j}(\omega, \omega',\tilde \mu) =  \left[
\begin{array}{ccccccc}
0 & 0  & 0 & 0 & 0 & 0 & 0  \\
0 & 0  & 0 & 0 & 0 & 0 & 0  \\
0 & 0 & 0  & 0 & 0 & \gamma_6^{\mu \, \sigma}(\omega, \omega') & 0  \\
0 & 0  & 0 & 0 & 0 & 0 & 0  \\
0 & 0  & 0 & 0 & 0 & 0 & 0  \\
0 & 0  & 0 & 0 & 0 & 0 & 0  \\
0 & 0  & 0 & 0 & 0 & 0 & 0  
\end{array}
\right]. \nn
\end{eqnarray}
With the following entries in the anomalous dimension matrix,
\begin{eqnarray}
\gamma_1(\omega, \omega',\tilde\mu)&=&- \frac{\alpha_s(\tilde\mu)}{\pi}C_F\left[
\delta(\omega-\omega')
+\frac{2}{\tilde\mu}
\phi_0'\left(\frac{\omega-\omega'}{\tilde\mu}\right)
\right], \nonumber \\
\gamma_2(\omega, \omega', \tilde\mu) &=& -2 \alpha_s(\tilde\mu) \frac{\omega'  \, C_F }{\pi \, m_b(\tilde\mu) }\delta(\omega-\omega'), \nonumber\\
\gamma_3^{\mu}(\omega, \omega', \tilde\mu) &=& -3 \alpha_s(\tilde\mu) \frac{ C_F }{2 \, \pi}\delta(\omega-\omega') \, v^{\mu}, \nonumber\\
\gamma_4(\omega, \omega', \tilde\mu) &=& \gamma_1(\omega, \omega',\tilde\mu)  + \frac{ \alpha_s(\tilde \mu) \, C_A}{ 2 \, \pi }\delta(\omega-\omega'), \nonumber\\
\gamma_5(\omega, \omega', \tilde\mu) &=& - \frac{\alpha_s(\tilde \mu) \, C_A}{ 2 \, \pi}\delta(\omega-\omega'),\nonumber \\
\gamma_6^{\mu \, \sigma}(\omega, \omega', \tilde\mu) &=& \frac{\alpha_s(\tilde \mu) \, C_A}{4 \, \pi}  ( i \, \epsilon^{\mu \, \sigma}_{\perp}) \delta(\omega-\omega'), \nonumber\\
\gamma_7^{\mu \, \sigma \, \eta}(\omega, \omega',\tilde \mu) &=& - \frac{\alpha_s(\tilde \mu) \, C_A }{4 \, \pi} ((v^\sigma  - n^\sigma) g^{\mu \, \eta} + n^{\eta} ( g^{\sigma \mu} - v^\sigma \, v^\mu)) \delta(\omega-\omega').
\end{eqnarray}

\section{Conclusions} \label{conclusions}

We have examined the anomalous dimension matrix appropriate for the phase space restricted  \Budecay and  \Bsgamma  \, decay spectra to subleading nonperturbative order. 
The effects of the time ordered products of the HQET Lagrangian with the leading order shape function operator were determined and the renormalizability and closure of a subset of the non-local operator basis used to describe these spectra, to subleading order, was established.

Operator mixing was found between the operators which occur to subleading order, requiring that the subleading operator basis be extended to include the operator $\bar{Q}_1$. This requires the introduction of  new shape functions to characterize the decay spectra of \Budecay and  \Bsgamma beyond tree level. The mixing determined between the operators $Q_1$ and $\bar{Q}_1$ is of the pernicious form that required a one gluon external state calculation to determine, despite the non vanishing zero gluon Feynman rules of the operators. We have also demonstrated that the possible mixing with the operator $Q^{\mu, \nu}(\omega_1, \omega_2, \Gamma)$ in a similar manner;
with vanishing Feynman rules for zero and one gluon, requires a two gluon external state calculation to completely determine the anomalous dimension at subleading non perturbative order.

Mixing was also determined between the  T product $T_{(O_0,O_k)}$ and the leading order shape function, and the T product $T_{(O_0,O_m)} $ was shown to lead to mixing between the $P_i$ and $O_i$ operators at this order.

The anomalous dimension and running of the $ \bar{Q}_1^\mu, Q_2^\mu $ and $ Q_4 $ operators was shown to be identical to the leading order shape function $Q_0$.

This work can be built upon in a number of ways. The anomalous dimension of the operator $Q_3$ is under investigation by the authors to establish the closure at one loop of the  set of subleading operators discussed in this paper. The anomalous dimension of the subleading four quark operators should be investigated to determine the full anomalous dimension matrix at subleading order. Once the full anomalous dimension is determined, Sudakov logarithms in the perturbative corrections to the subleading operators can be resummed, so that renormalization group improved calculations can be undertaken for the  \Budecay and  \Bsgamma \, decay spectra to subleading nonperturbative order.

More important than the small effect that these corrections have directly on the extraction of $|V_{ub}|$, is the fact that this work establishes the renormalizability of a subset of the soft sector nonperturbative expansion beyond leading order. This is a necessary step in extending QCD factorization theorems beyond leading nonperturbative order, validating the factorization based approach used for the phase space restricted  $\btou$ and \Bsgamma \, decay spectra beyond leading nonperturbative order.


\section{Acknowledgments}
\label{acknowledgments}

This work was begun in with the initial involvement of M. Luke, and we thank him for his support 
and guidance throughout.
Financial support was provided by the Natural Sciences and Engineering
Research Council of Canada. M. Trott also acknowledges the support of the 
Walter B. Sumner Foundation. 
A. Williamson acknowledges financial support by the Director, Office of Science, Office of 
High Energy, Division of High Energy Physics under Contract 
DOE-ER-40682-143 and DEAC02-6CH03000.

\end{document}